\documentclass[aps, preprint, groupedaddress, floatfix]{revtex4}

\usepackage{epsfig}
\usepackage{epstopdf}
\usepackage{graphicx}
\usepackage{tabularx}
\usepackage{multirow}
\usepackage{amsmath}
\usepackage{color}

\newcolumntype{Y}{>{\centering\arraybackslash}X}
\newcolumntype{Z}{>{\centering\arraybackslash}X}

\begin{document}

\title{Phase diagram of Sr$_{1-x}$Ba$_x$MnO$_3$ as a function of chemical doping, 
epitaxial strain and external pressure}

\author{Hanghui Chen$^{1,2}$ and Andrew J. Millis$^{1}$}

\affiliation{
 $^1$Department of Physics, Columbia University, New York, NY, 10027, USA\\
 $^2$Department of Applied Physics and Applied Mathematics, Columbia University, New York, NY, 10027, USA\\}
\date{\today}

\begin{abstract}
  We use \textit{ab initio} calculations to systematically study the
  phase diagram of multiferroic Sr$_{1-x}$Ba$_x$MnO$_3$
  ($0 \leq x \leq 1$) as a function of chemical doping, epitaxial
  strain and external pressure. We find that by replacing Sr with Ba
  in cubic SrMnO$_3$ and imposing epitaxial strain, the material can
  be tuned to the vicinity of a first order transition between two
  multiferroic phases, one antiferromagnetic with a smaller
  polarization and one ferromagnetic with a larger polarization. A
  giant effective magneto-electric coupling and cross-field control
  (electric field control of magnetism or magnetic field control of
  polarization) can be achieved in the vicinity of the transition. The
  dependence of the theoretically computed transition point on the
  choice of exchange correlation functionals is determined and is
  found to be non-negligible. We also show that the perovskite
  structure of BaMnO$_3$ can be stabilized relative to its
  hexagonal polymorphs at pressures larger than 20 GPa.
\end{abstract}

\maketitle

\section{Introduction}

Searching for multiferroic materials, in particular those
simultaneously possessing ferroelectric polarization and ferromagnetic
moments with strong coupling between the two order parameters, is
currently of great
interest~\cite{Eerenstein-Nature-2006,Ramesh-NatMat-2007}. 
Among all single-phase multiferroic materials, perovskite
Sr$_{1-x}$Ba$_x$MnO$_3$ ($0 \leq x \leq 1$) stands out as a promising
candidate for multiferroics in theory~\cite{Hong-PRB-2012,
  Sakai-PRL-2011, Giovannetti-PRL-2012,
  Nourafkan-PRB-2014, Langenberg-ACSAMI-2015, Pratt-PRB-2014}. 
Ferroelectricity was theoretically predicted in
perovskite BaMnO$_3$~\cite{Rondinelli-PRB-2009}.  It was also
predicted~\cite{Lee-PRL-2010, Fennie-PRL-2006} that under biaxial strain, perovskite
SrMnO$_3$ (a paraelectric antiferromagnet at ambient conditions)
becomes a ferroelectric ferromagnet.

In this paper, we examine the phase diagram of
Sr$_{1-x}$Ba$_{x}$MnO$_3$ ($0 \leq x\leq 1$) as a function of chemical
doping, epitaxial strain and external pressure.  We show that by
doping cubic SrMnO$_3$ with an appropriate concentration of Ba and
imposing epitaxial strain, the material can be tuned to the vicinity
of a first order phase transition between two multiferroic phases, one
antiferromagnetic with a smaller polarization and one ferromagnetic
with a larger polarization. In the vicinity of the transition point,
reasonable electric fields can be used to switch the material between
ferromagnetic and antiferromagnetic states while feasible magnetic
fields can change the electrical polarization.  However, the critical
Ba doping and the critical epitaxial strain strongly depend on the
exchange correlation functionals. Building on our previous
work~\cite{Chen-PRB-2016c}, we compare three different versions of
Perdew-Burke-Ernzerhof (PBE) parameterized exchange correlation
functionals and show that they make substantially different
predictions for the position of the magnetic transition boundary in
the phase diagram of Ba doping and epitaxial strain. We focus on the
perovskite structure of Sr$_{1-x}$Ba$_{x}$MnO$_3$ to achieve effective
magneto-electric couplings. However, BaMnO$_3$ can also crystallize in
hexagonal polymorph, which is more stable than the perovskite
structure. We show that applying an external pressure of over 20 GPa
on BaMnO$_3$ is a feasible approach to stabilize the more
interesting perovskite structure over its hexagonal polymorph.

The rest of this paper is organized as follows. Sec.~\ref{sec:com}
presents the computational details. Sec.~\ref{sec:chem} discusses
the phase diagram of perovskite Sr$_{1-x}$Ba$_{x}$MnO$_3$ 
as a function of chemical doping and estimates the fields required for magneto-electric
cross-field control. Sec.~\ref{sec:strain} presents the magnetic phase diagram of perovskite
SrMnO$_3$, Sr$_{0.5}$Ba$_{0.5}$MnO$_3$ and BaMnO$_3$ as a function of
strain. Sec.~\ref{sec:pressure} studies energetic dependence of 
perovskite and hexagonal polymorphs of BaMnO$_3$ on external pressure.
We conclude in Sec.~\ref{sec:conclusion}.

\section{\label{sec:com} Computational Details}

We perform density functional theory
calculations~\cite{Hohenberg-PR-1964,Kohn-PR-1965} within the
~\textit{ab initio} plane-wave approach~\cite{Payne-RMP-1992}, as
implemented in the Vienna Ab-initio Simulation Package
(VASP)~\cite{Kresse-PRB-1996}. We employ projector augmented wave
(PAW) pseudopotentials~\cite{Blochl-PRB-1994, Kresse-PRB-1999}. We use
an energy cutoff 600 eV. All the calculations allow for
spin-polarization to study different types of long-range magnetic
orderings. For most calculations, both cell and internal coordinates
are fully relaxed until each force component is smaller than 10
meV/\AA~and stress tensor is smaller than 10 kBar. For the study of
strain, the in-plane lattice constants are fixed during atomic
relaxation; the internal coordinates and out-of-plane lattice constant
are fully relaxed.  For the study of pressure, the atomic relaxation
is converged to a specific external pressure.

We consider different structures of Sr$_{1-x}$Ba$_x$MnO$_3$.  
A $12\times 12\times 10$ Monkhorst-Pack grid is used to
sample the Brillouin zone of the simulation cell used in 
Sec.~\ref{sec:chem} and Sec.~\ref{sec:strain} to study the perovskite structure.
For the hexagonal polymorphs studied in Sec.~\ref{sec:pressure}, 
a $10\times 10\times 10$ Monkhorst-Pack grid is used.  
The simulation cells are provided in each
section for ease of reading.

Our recent study shows~\cite{Chen-PRB-2016c} that different exchange correlation functionals
make substantially different predictions for structural and magnetic
properties of ferroelectric manganites.  By
comparing to available experimental data for
Sr$_{0.5}$Ba$_{0.5}$MnO$_3$ alloy, we establish that
charge-density-only generalized gradient approximation with
Perdew-Burke-Ernzerhof parameterization~\cite{Perdew-PRL-1996} plus Hubbard $U$ and Hund's
$J$ corrections (PBE+$U$+$J$)~\cite{Lie-PRB-1995} and spin-polarized generalized gradient
approximation with Perdew-Burke-Ernzerhof parameterization revised for
solids (sPBEsol)~\cite{Perdew-PRL-2008} make overall better predictions than the
spin-polarized PBE (sPBE) and local spin density approximation (LSDA).
In this paper, for most of the results, we use PBE+$U$+$J$ and
sPBEsol. For reference, we also compare PBE+$U$+$J$ and sPBEsol to
sPBE plus effective Hubbard $U_{\rm eff}$ corrections.  For
the PBE+$U$+$J$ method, we use accepted values of $U$ = 5 eV and $J$ = 0.7
eV for Mn $d$ orbitals unless otherwise specified. This method is
implemented in VASP by turning on LDAUTYPE = 4. For the sPBE+$U_{\rm eff}$
method, we follow previous work~\cite{Lee-PRL-2010} and use
$U_{\rm eff}$ = 1.7 eV. This method is
implemented in VASP by turning on LDAUTYPE = 1. 

\section{\label{sec:chem} Chemical doping}

\begin{figure}[t]
\includegraphics[angle=0,width=0.85\textwidth]{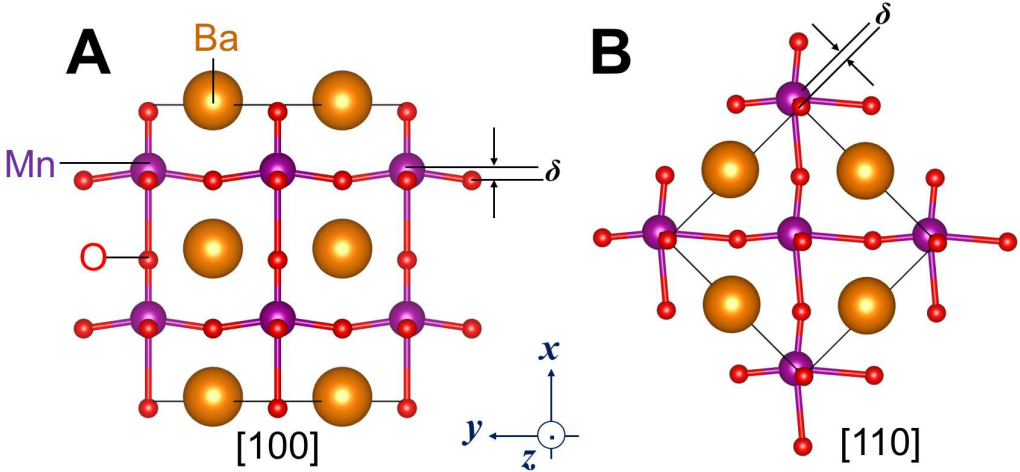}
\caption{\label{fig:BMOstructure} Atomic structure of perovskite BaMnO$_3$.
The orange, purple and red balls are
Ba, Mn and O atoms, respectively. The Mn off-center displacement 
$\vec{\delta}_{\textrm{Mn-O}}$ is formally defined as:
$\vec{\delta}_{\textrm{Mn-O}} = \frac{1}{6}\sum^{6}_{i=1}\vec{R}_{\textrm{Mn-O}}$
where $\vec{R}_{\textrm{Mn-O}}$ are the vectors connecting a Mn ion 
to its six nearest oxygen neighbors. \textbf{A}) Mn off-center
displacement $\delta$ along the [100] direction. \textbf{B}) Mn off-center
displacement $\delta$ along the [110] direction.}
\end{figure}

\begin{figure}[t]
\hspace*{-0.6cm}\includegraphics[angle=0,width=0.95\columnwidth]{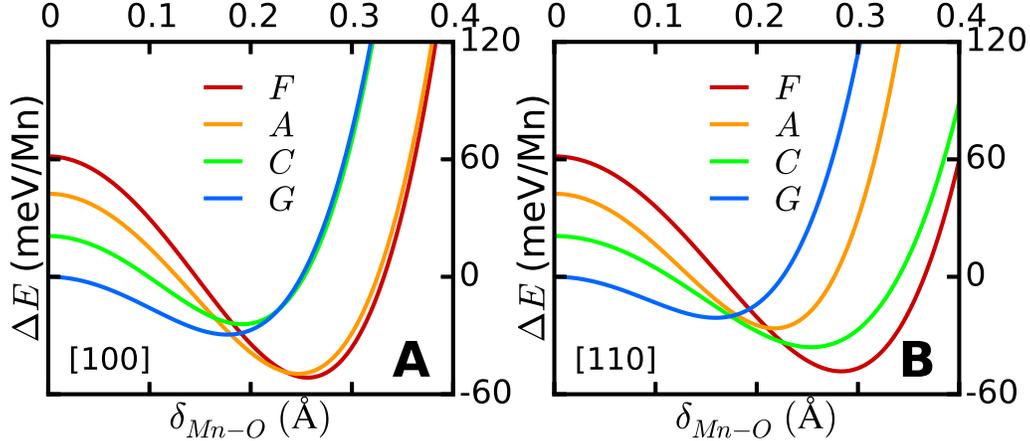}
\caption{\label{fig:double} Change in total energy of perovskite
  BaMnO$_3$ as a function of Mn off-center displacement
  $\delta_{\textrm{Mn-O}}$ computed using the PBE+$U$+$J$ method with $U$ = 5 eV and
  $J$ = 0.7 eV for ferromagnetic ($F$, red curves), $A$-type
  antiferromagnetic ($A$, orange curves), $C$-type antiferromagnetic
  ($C$, green curves) and $G$-type antiferromagnetic ($G$, blue
  curves) states (for the definition of magnetic orderings, see the
  main text) for \textbf{A}) along the [100] and \textbf{B}) along the
  [110] directions of $\delta_{\textrm{Mn-O}}$. The zero energy state
  is chosen as the energy of the $G$-type antiferromagnetic state at
  $\delta_{\textrm{Mn-O}}$ = 0. Calculations were first performed
  using a cubic structure ($\delta_{\textrm{Mn-O}}=0$) and a fully
  relaxed structure, then were interpolated by using intermediate
  structures.}
\end{figure}

Following our previous work~\cite{Chen-PRB-2016c}, we first use the PBE+$U$+$J$ method to
examine the Ba-end member compound perovskite BaMnO$_3$, considering four
different magnetic orderings that are commonly found in manganites:
ferromagnetic ($F$), two-sublattice N\'{e}el ordering ($G$-type), (1,
0, 0) stripe ordering ($A$-type) and (1, 1, 0) stripe ordering
($C$-type). Fig.~\ref{fig:BMOstructure} shows the atomic structure 
of perovskite BaMnO$_3$. We use a $\sqrt{2}\times\sqrt{2}\times 2$ 
simulation cell that can accommodate all the above four 
magnetic orderings. In Fig.~\ref{fig:BMOstructure}, $\delta$ 
is the Mn off-center displacement which is along the 
[100] axis (Fig.~\ref{fig:BMOstructure}\textbf{A}) and along 
the [110] axis (Fig.~\ref{fig:BMOstructure}\textbf{B}).

Fig.~\ref{fig:double} presents the dependence of the 
total energy on the Mn off-center displacement
$\delta_{\textrm{Mn-O}}$ for both [100] and [110] directions. The
reference energy is that of the $G$-type antiferromagnet at
$\delta_{\textrm{Mn-O}}=0$. All the other magnetic states
exhibit an energy minimum at $\delta_{\textrm{Mn-O}}\neq 0$, but the
ferromagnetic state is lowest for both directions, and has the largest
$\delta_{\textrm{Mn-O}}$ at the minimum. Interestingly, the energy
difference between different magnetic phases depends on the
direction of polarization. For $\delta_{\textrm{Mn-O}}$ along the
[110] direction, ferromagnetism is clearly more stable than any of the
antiferromagnetic states, while for $\delta_{\textrm{Mn-O}}$ along the
[100] direction, ferromagnetism and $A$-type antiferromagnetism are
extremely close in energy. The near coincidence of energies suggests
that for $x$ slightly less than $1$, it may be possible to switch the
magnetic states by applying an electric field and rotating the
orientation of $\delta_{\textrm{Mn-O}}$.


\begin{figure}[t]
\includegraphics[angle=0,width=\columnwidth]{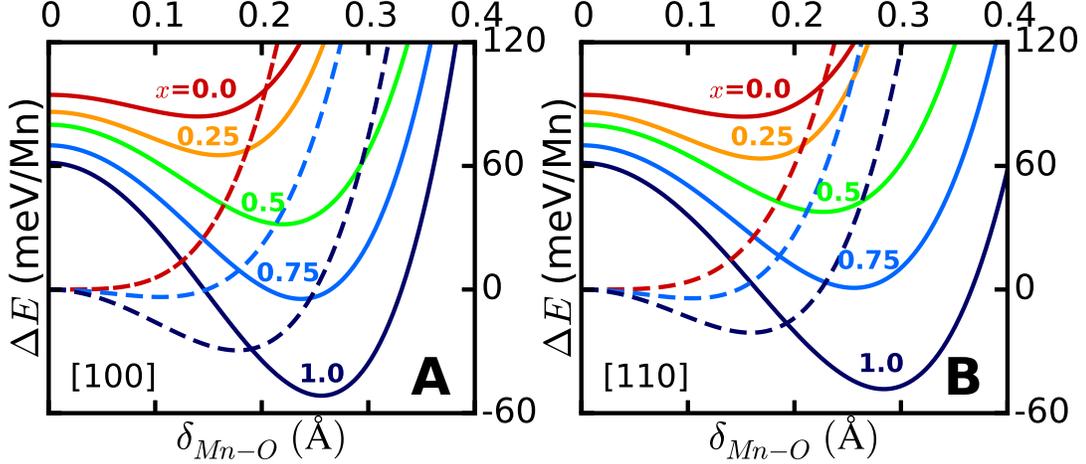}
\caption{\label{fig:landau} Total energy of Sr$_{1-x}$Ba$_x$MnO$_3$
  ($0 \leq x \leq 1$) calculated using the PBE+$U$+$J$ method with $U$ = 5 eV and
  $J$ = 0.7 eV as a function of Mn off-center displacement
  $\delta_{\textrm{Mn-O}}$. The solid lines indicate ferromagnetic
  ordering and the dashed lines indicate $G$-type antiferromagnetic
  ordering. The red, orange, green, blue and indigo curves correspond to 
  Ba doping of $x$ = 0, 0.25, 0.5, 0.75 and 1, respectively. 
  For each $x$, the zero of energy is chosen as the energy
  of the $G$-type antiferromagnet at $\delta_{\textrm{Mn-O}} = 0$.
  \textbf{A}) $\delta_{\textrm{Mn-O}}$ along the [100] direction;
  \textbf{B}) $\delta_{\textrm{Mn-O}}$ along the [110] direction.}
\end{figure}

\begin{figure}[t]
\includegraphics[angle=0,width=\columnwidth]{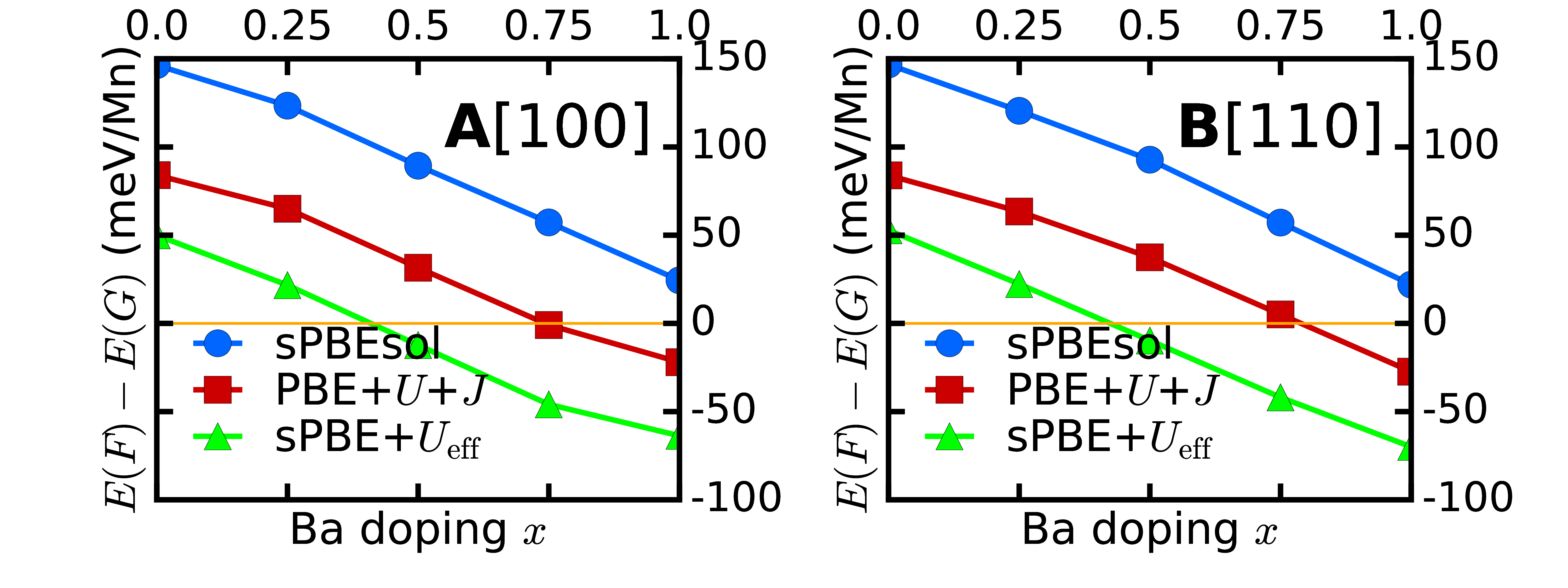}
\caption{\label{fig:energydiffdoping} Energy difference between
  ferromagnetic state ($F$) and $G$-type antiferromagnetic state ($G$)
  of Sr$_{1-x}$Ba$_x$MnO$_3$
  ($0 \leq x \leq 1$) calculated using difference exchange correlation functionals:
  charge-density-only PBE functional with Hubbard $U$ and Hund's $J$
  correction (PBE+$U$+$J$) with $U$ = 5 eV and
  $J$ = 0.7 eV; spin polarized PBE for solids (sPBEsol) and 
  spin polarized PBE with effective Hubbard $U_{\rm eff}$ correction (sPBE+$U_{\rm eff}$).
  \textbf{A}) $\delta_{\textrm{Mn-O}}$ along the [100] direction;
  \textbf{B}) $\delta_{\textrm{Mn-O}}$ along the [110] direction.}
\end{figure}

The results shown in Fig.~\ref{fig:landau} establish that within the
PBE+$U$+$J$ method with $U$ = 5 eV and $J$ = 0.7 eV, perovskite BaMnO$_3$ is a
ferromagnet with strong Mn off-center displacements. Now we consider
the evolution of magnetic and ferroelectric states with Ba
concentration $x$ in Sr$_{1-x}$Ba$_x$MnO$_3$.  Fig.~\ref{fig:landau} presents the
dependence of the total energy of two magnetic states of
Sr$_{1-x}$Ba$_x$MnO$_3$ (for $x$ = 0.0, 0.25, 0.5, 0.75 and 1.0) on the
inversion-symmetry-breaking Mn-O displacement $\delta_{\textrm{Mn-O}}$
along [100] and [110] directions. The figure shows that for both
directions of $\delta_{\textrm{Mn-O}}$, the $G$-type antiferromagnetic
state (which is insulating for all $x$) has a
paraelectric-to-ferroelectric transition as the Ba doping $x$
increases, consistent with previous studies~\cite{Rondinelli-PRB-2009,
  Sakai-PRL-2011,Giovannetti-PRL-2012, Nourafkan-PRB-2014}. We also
see that for all $x$ the ferromagnetic state always has a nonzero Mn
off-center displacement $\delta_{\textrm{Mn-O}}$ which is larger than
that of antiferromagnetic state at the same $x$. With increasing Ba
doping the energy difference between the ferromagnetic and
antiferromagnetic states rapidly decreases and for the parameters $U$
= 5 eV and $J$ = 0.7 eV used here, a first-order
antiferromagnetic-to-ferromagnetic transition occurs around $x=0.75$
(blue curves), accompanied by an abrupt jump in the Mn off-center
displacement. Therefore, in the vicinity of the transition, we can use
an external electric field to increase the Mn off-center displacement
and change the magnetic ordering, or use an external magnetic field to
induce a finite change in Mn off-center displacements. This effective
magneto-electric cross-field control is similar to what is found in
epitaxially strained SrMnO$_3$, indicating that chemical doping is a
complementary approach to tuning multiferroic properties.


Based on Fig.~\ref{fig:landau}, we can make some estimation of critical
fields required for switching. For physically reasonable
parameters $U$ = 5 eV and $J$ = 0.7 eV, our calculations show that
the critical doping $x$ is around 0.75. At $x = 0.75$ and along
the [110] direction, the two
lowest energy states are $G$-type antiferromagnetism ($G$) and
ferromagnetism ($F$). The critical fields are estimated by:

\begin{equation}
\label{eqn:efield} \mathcal{E}_c\cdot(P_F - P_G) \simeq E(F) -
E(G)
\end{equation}

\begin{equation}
\label{eqn:bfield} \mathcal{H}_c\cdot M \simeq E(F) - E(G)
\end{equation}
where $P_F$ = 56 $\mu$C/cm$^2$ and $P_G$ = 26 $\mu$C/cm$^2$ are 
electric polarizations calculated for ferromagnetic and $G$-type
antiferromagnetic states with on-site magnetic moment $M\simeq 3\mu_B$
using the Berry phase method~\cite{King-Smith-PRB-1993} (their
electric field dependence is neglected). We find that $\mathcal{E}_c$
is about 42 MV/m and $\mathcal{H}_c$ is about 25 T. While these fields
are large, they are experimentally achievable and have been applied
to, for example, BaTiO$_3$ thin films~\cite{Jo-APL-2006}. More
importantly, tuning the Ba doping $x$ by a few percent can rapidly
reduce these critical fields. Use of other interaction parameters $U$ = 4 eV
and $J$ = 0.6 eV (the smallest $U$ value to stabilize a high-spin state
in cubic SrMnO$_3$ using the PBE+$U$+$J$ method~\cite{Chen-PRB-2016}) 
shows that at $x$ = 1, we can still switch the
two lowest magnetic states by a critical magnetic field of 12 T or
switch those two polarizations by an electric field of 30 MV/m. As we
see, properly adjusting the Ba doping $x$ can compensate for the
uncertainty of $U$ and $J$ parameters in theory.

Our previous works~\cite{Chen-PRB-2016, Chen-PRB-2016c} show that magnetic properties of transition
metal oxides depend on the choice of exchange-correlation
functionals. To test whether our prediction of a doping-driven magnetic
transition in Sr$_{1-x}$Ba$_x$MnO$_3$ is robust to the choice of
exchange-correlation potential, we compare three different types
of PBE exchange correlation functionals: spin-polarized PBE revised
for solids (sPBEsol), charge-density-only PBE plus Hubbard $U$ and Hund's $J$
correction (PBE+$U$+$J$) and spin-polarized PBE plus effective
$U_{\rm eff}$ correction (sPBE+$U_{\rm eff}$) by calculating the
energy difference of Sr$_{1-x}$Ba$_x$MnO$_3$ between the ferromagnetic
and $G$-type antiferromagnetic states. The comparison is shown in
Fig.~\ref{fig:energydiffdoping}.  We find that while PBE+$U$+$J$ and
sPBE+$U_{\rm eff}$ predict that there is an antiferromagnetic to
ferromagnetic transition as $x$ is increased from 0 to 1 for 
perovskite Sr$_{1-x}$Ba$_x$MnO$_3$, 
sPBEsol predicts that perovskite Sr$_{1-x}$Ba$_x$MnO$_3$ is $G$-type
antiferromagnetic for the whole range of $x$ ($0\leq x \leq 1$).

We can test the functionals by comparing to experimental data for
Sr$_{0.5}$Ba$_{0.5}$MnO$_3$ alloy~\cite{Sakai-PRL-2011}. We find from
our previous study~\cite{Chen-PRB-2016c} that both
PBE+$U$+$J$ and sPBEsol make good predictions of structural and
magnetic properties (better than sPBE and
LSDA). Hence we can not definitely predict the
ground state magnetic ordering of perovskite BaMnO$_3$. However, that
doping cubic SrMnO$_3$ with Ba favors ferromagnetism is a robust
result. If the end member perovskite BaMnO$_3$ were ferromagnetic,
then the antiferromagnetic-to-ferromagnetic transition would occur at
an intermediate doping $x_c$, in the vicinity of which an effective
giant magneto-electric coupling can be induced. If perovskite
BaMnO$_3$ were not ferromagnetic, then epitaxial strain may be used to
stabilize the ferromagnetic state.

\section{\label{sec:strain} Epitaxial strain}


Strain engineering has been widely used to induce multiferroic
states~\cite{Fennie-PRL-2006, Lee-Nature-2010}. Lee and Rabe
showed~\cite{Lee-PRL-2010} that epitaxial strain can induce a
ferroelectric-ferromagnetic ground state in an otherwise
antiferromagnetic cubic SrMnO$_3$ with a critical tensile strain of
3.4\% (predicted by sPBE plus an effective Hubbard $U_{\rm eff}$
extension, for short sPBE+$U_{\rm eff}$).  In this section, we define
the strain $\lambda$ as:

\begin{equation}
\label{eqn:lambda}\lambda=\frac{a-a_0}{a_0}\times 100\%
\end{equation}
where $a$ is a common lattice constant that is imposed on all the
magnetic orderings and $a_0$ is the theoretically calculated lattice
constant of the $G$-type antiferromagnetic state. We note that because
we define strain with respect to a particular state (the $G$-type
antiferromagnet), the critical strain depends both on the energy
difference between ferromagnetic and antiferromagnetic states at
equilibrium (denoted by $\delta E$) and on the difference between
equilibrium lattice constant (denoted by $\delta \lambda$).  We first
compare the predictions of sPBE+$U_{\rm eff}$ to PBE+$U$+$J$ and
sPBEsol methods for the phase diagram of strained SrMnO$_3$. We show
that the three methods make substantially different predictions on the
critical strain that stabilizes the ferroelectric-ferromagnetic ground
state of SrMnO$_3$.

Next we study the phase diagram of epitaxially strained perovskite
Sr$_{0.5}$Ba$_{0.5}$MnO$_3$ alloy and epitaxially strained perovskite
BaMnO$_3$.  We show that Ba doping acts like an ``effective
strain'', so that increasing the Ba concentration in
Sr$_{1-x}$Ba$_x$MnO$_3$ can reduce the critical strain that stabilizes
the ferromagnetic state. However, just like epitaxially strained
SrMnO$_3$, the critical strain as a function of Ba doping also strongly
depends on exchange correlation functionals.

\subsection{Perovskite SrMnO$_3$}

\begin{figure}[t]
\includegraphics[angle=0,width=0.85\textwidth]{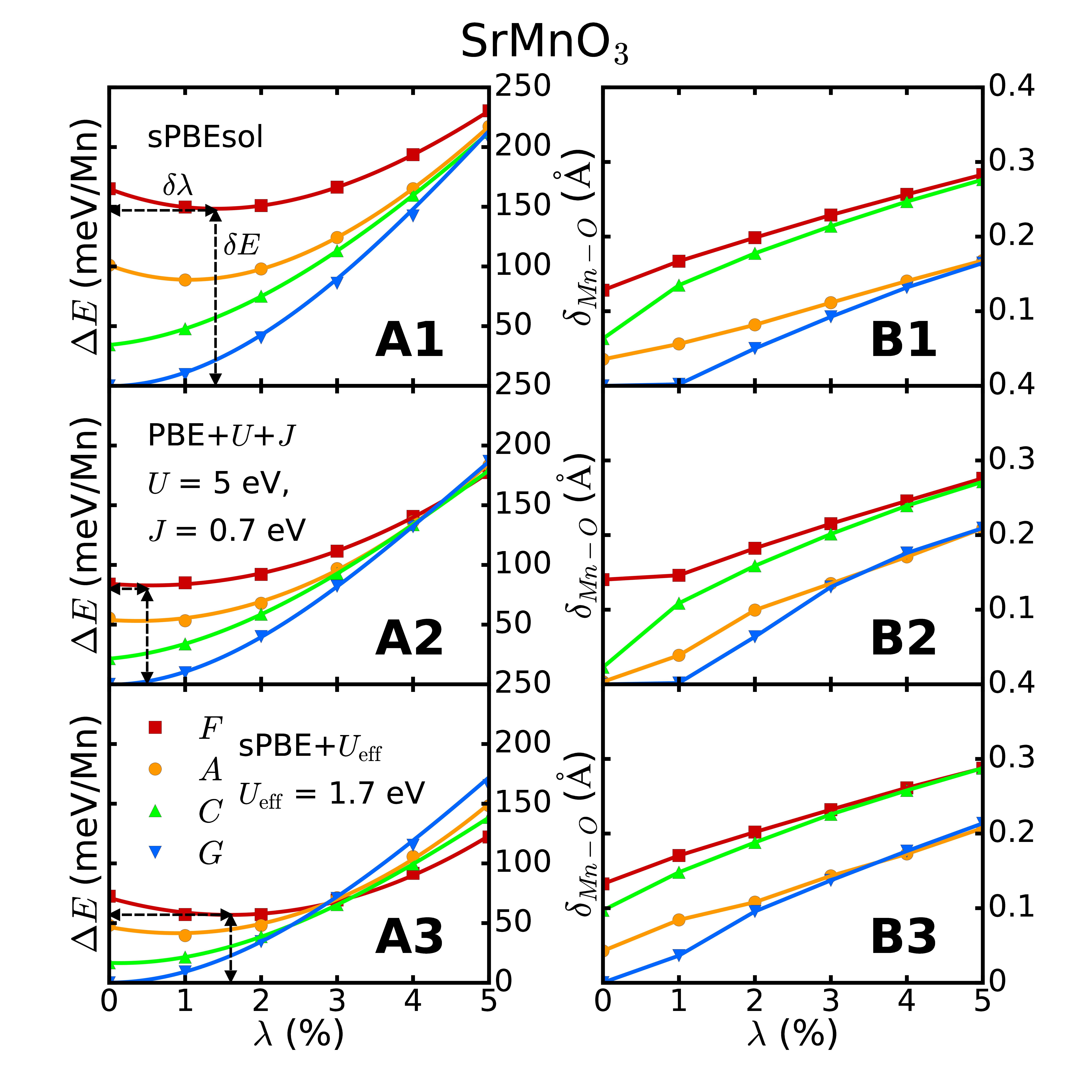}
\caption{\label{fig:strainSMO} Perovskite SrMnO$_3$: \textbf{A1}) and
  \textbf{B1}) using the sPBEsol method.  \textbf{A1}) Total energy
  for different magnetic states under tensile strain. The
  zero point is chosen as the total energy of $G$-type
  antiferromagnetic state at zero strain. $\lambda$ is the strain with
  respect to the equilibrium lattice constant of $G$-type
  antiferromagnetic state. \textbf{B1}) Mn off-center displacement
  along the [110] direction of SrMnO$_3$ for different magnetic states
  under tensile strain. \textbf{A2}) and \textbf{B2}) using the
  PBE+$U$+$J$ method with $U$ = 5 eV and $J$ = 0.7 eV.
  \textbf{A2}) same as \textbf{A1}), \textbf{B2}) same as
  \textbf{B1}). \textbf{A3}) and \textbf{B3}) using the sPBE+$U_{\rm eff}$ method
  with $U_{\rm eff}$ = $U-J$ = 1.7 eV. \textbf{A3}) same as
  \textbf{A1}), \textbf{B3}) same as \textbf{B1}). The red squares
  and lines are the ferromagnetic state ($F$). The orange squares and
  lines are the $A$-type antiferromagnetic state ($A$). The green
  squares and lines are the $C$-type antiferromagnetic state ($C$). The
  blue squares and lines are the $G$-type antiferromagnetic state
  ($G$). The horizontal dashed line denotes $\delta \lambda$, which is the difference in equilibrium lattice
  constants between $F$ and $G$ orderings. The vertical dashed line
  denotes $\delta E$, which is the equilibrium energy
  difference between $F$ and $G$ orderings. }
\end{figure}

\begin{figure}[t]
\includegraphics[angle=0,width=\textwidth]{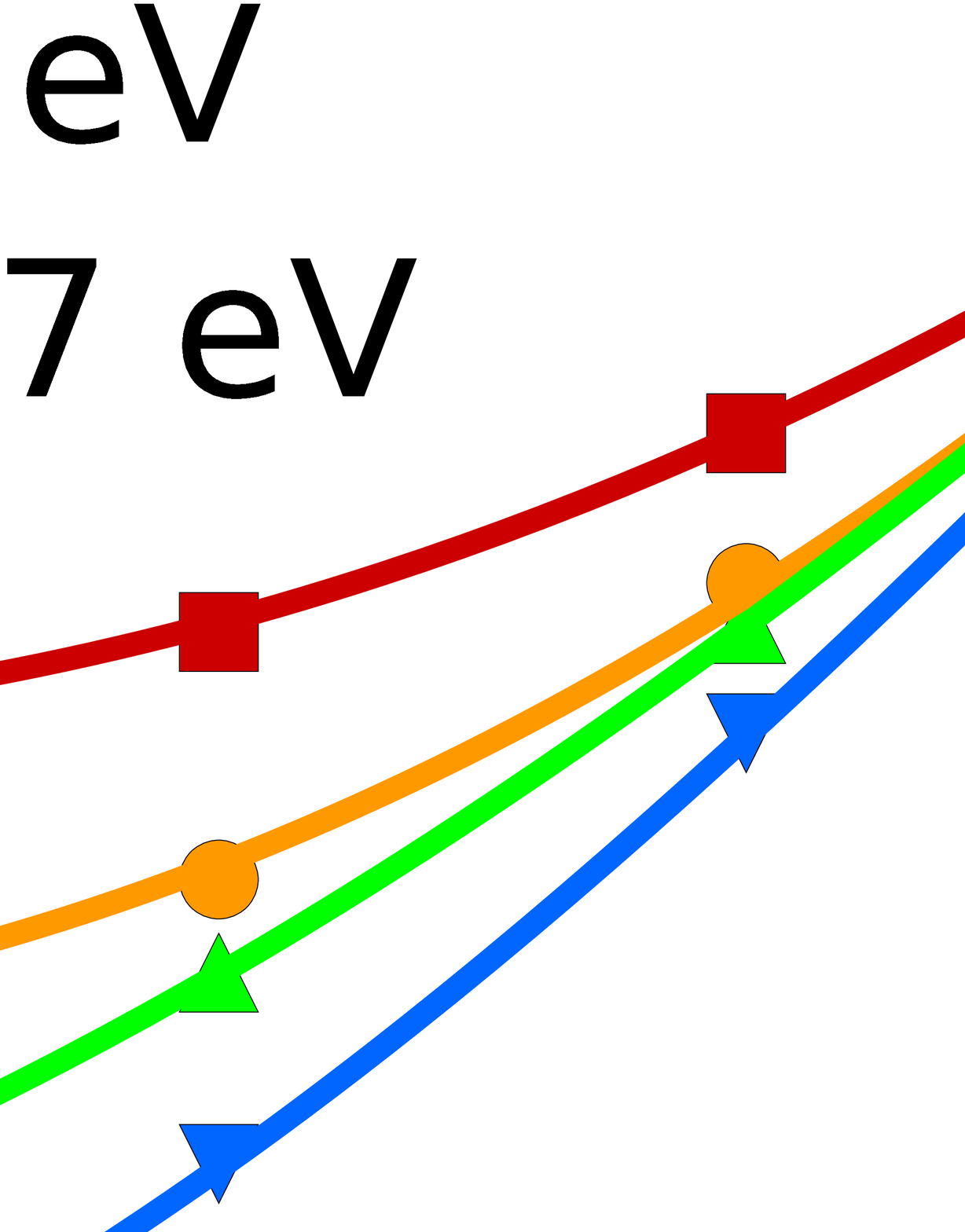}
\caption{\label{fig:strainSMOU} Calculations of total energies of SrMnO$_3$
  for different magnetic states under tensile strain, using the
  PBE+$U$+$J$ method (\textbf{A1} $U$ = 4 eV; \textbf{A2} $U$ = 5 eV;
  \textbf{A3} $U$ = 6 eV; for all three cases, $J$ = 0.7 eV).
  The red squares
  and lines are the ferromagnetic state ($F$). The orange squares and
  lines are the $A$-type antiferromagnetic state ($A$). The green
  squares and lines are the $C$-type antiferromagnetic state ($C$). The
  blue squares and lines are the $G$-type antiferromagnetic state
  ($G$). The horizontal dashed line denotes $\delta \lambda$, which is the difference in equilibrium lattice
  constants between $F$ and $G$ orderings. The vertical dashed line
  denotes $\delta E$, which is the 
  equilibrium energy difference between $F$ and $G$ orderings.}
\end{figure}

Fig.~\ref{fig:strainSMO} shows the total energies of SrMnO$_3$ and Mn
off-center displacement along the [100] direction under tensile
strain.  Fig.~\ref{fig:strainSMO}\textbf{A1} and \textbf{B1} are
calculated using the sPBEsol method.
Fig.~\ref{fig:strainSMO}\textbf{A2} and \textbf{B2} are calculated
using the PBE+$U$+$J$ method with $U$ = 5 eV and $J$ = 0.7 eV.
Fig.~\ref{fig:strainSMO}\textbf{A3} and \textbf{B3} are calculated
using the sPBE+$U_{\rm eff}$ method with $U_{\rm eff}$ = $U-J$ = 1.7
eV, following Ref.~\cite{Lee-PRL-2010}.  The panels \textbf{A} of
Fig.~\ref{fig:strainSMO} show the total energy of tensile strained
SrMnO$_3$ with different magnetic orderings ($F$, $A$, $C$ and $G$, as
defined in the previous section).  We find the predicted critical tensile strain
needed to drive a ferromagnetic transition is 6.8\% for sPBEsol (not
explicitly shown in the Fig.~\ref{fig:strainSMO}\textbf{A1}), 4.8\%
for PBE+$U$+$J$ and 3.2\% for sPBE+$U_{\rm eff}$, respectively. The
critical strain for sPBE+$U_{\rm eff}$ is slightly smaller than that
given in Ref.~\cite{Lee-PRL-2010} (see~\footnote{The slight difference
  between our results and Ref.~\cite{Lee-PRL-2010} is because our
  critical strain is extracted from calculations of $Amm2$ symmetry
  with polarization along the [110] direction while in
  Ref.~\cite{Lee-PRL-2010} the critical strain is extracted from
  calculations of $Ima2$ symmetry.}). We notice that all the three
methods predict that with increasing tensile strain, the elastic
energy eventually stabilizes the ferromagnetic state over the $G$-type
antiferromagnetic state.  However, the strain dependence are
substantially different. The difference has two origins. One is the
energy difference $\delta E$ between the ferromagnetic and $G$-type
antiferromagnetic states at equilibrium, which is shown as the dashed
vertical lines in the panels \textbf{A} of Fig.~\ref{fig:strainSMO}. The
other is the difference in equilibrium lattice constants $\delta\lambda$ between the
ferromagnetic state and $G$-type antiferromagnetic state, which is
shown as the horizontal lines in the panels \textbf{A} of
Fig.~\ref{fig:strainSMO}.  $\delta E$ is more important.  Increasing
$\delta E$ substantially increases the critical strain. sPBEsol,
PBE+$U$+$J$ and sPBE+$U_{\textrm{eff}}$ predict $\delta E$ to be 149
meV/Mn, 83 meV/Mn and 57 meV/Mn, respectively.  For a fixed
$\delta E$, increasing $\delta \lambda$ (within a reasonable range)
can reduce the critical strain, because at the same constrained
lattice constant, a larger $\delta \lambda$ means that the
ferromagnetic state is under smaller strain, thus inducing smaller
elastic energy cost. The sPBEsol, PBE+$U$+$J$ and sPBE+$U_{\rm eff}$
methods predict that the equilibrium lattice constant of the
ferromagnetic state is 1.4\%, 0.5\% and 1.6\% larger than that of the
$G$-type antiferromagnetic state. sPBE+$U_{\rm eff}$ predicts the
smallest $\delta E$ and the largest $\delta \lambda$, the combination of which
leads to the smallest critical strain ($3.4\%$). On the
other hand, sPBEsol predicts the largest $\delta E$, which
leads to the largest critical strain ($6.8\%$).

The panels \textbf{B} of Fig.~\ref{fig:strainSMO} show the Mn off-center
displacement $\delta_{\textrm{Mn-O}}$ along the [110] direction for
strained SrMnO$_3$. We find that all the three methods yield similar
behavior for $\delta_{\textrm{Mn-O}}$ with sPBE+$U_{\rm eff}$ predicting a
slightly larger displacement than sPBEsol and PBE+$U$+$J$, consistent with our
previous study~\cite{Chen-PRB-2016c}. 

Finally we briefly discuss the Hubbard $U$ dependence. We study the
$U$ dependence using the
PBE+$U$+$J$ method, but the conclusion also applies to sPBE+$U_{\rm eff}$.
We study several different values of Hubbard $U$ and compare the results in
Fig.~\ref{fig:strainSMOU}. As we increase the Hubbard $U$ from 4 to 5 to 6
eV, $\delta E$ (dashed vertical lines in Fig.~\ref{fig:strainSMOU}) is
reduced from 103 meV/Mn to 83 meV/Mn to 66 meV/Mn. At the same time, $\delta
\lambda$ (dashed horizontal lines in Fig.~\ref{fig:strainSMOU}) also
decreases from 1.2\%, to 0.5\% to 0.4\%. While a decreasing $\delta
E$ favors the ferromagnetism and a decreasing $\delta \lambda$ disfavors the
ferromagnetism, $\delta E$ is the dominating factor and eventually the
critical strain is reduced. Therefore, increasing $U$ favors the
ferromagnetism, which is consistent with the simple picture that a
larger $U$ suppresses the superexchange and antiferromagnetism, and
thus ferromagnetism is favored.

\subsection{Perovskite Sr$_{0.5}$Ba$_{0.5}$MnO$_3$ alloy}

\begin{figure}[t]
\includegraphics[angle=0,width=0.85\textwidth]{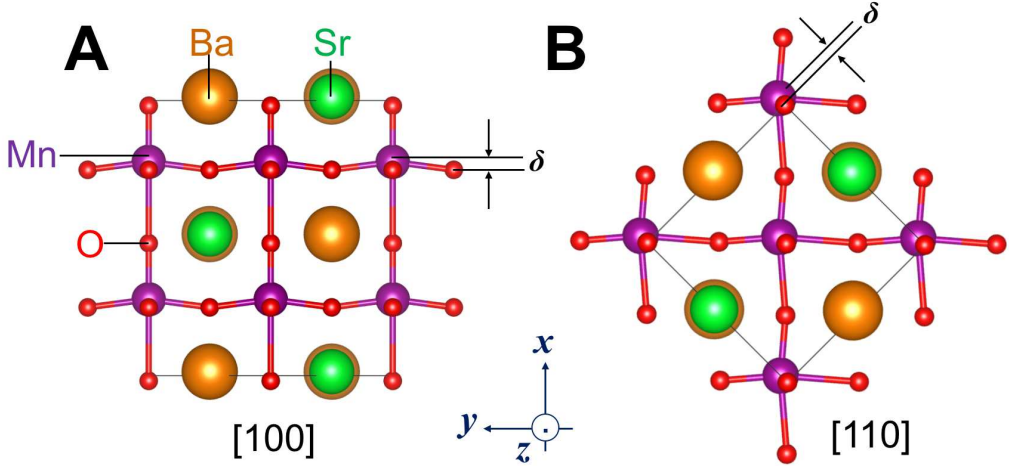}
\caption{\label{fig:SBMOstructure} Atomic structure of perovskite
Sr$_{0.5}$Ba$_{0.5}$MnO$_3$. The orange, green, purple and red balls are
Sr, Ba, Mn and O atoms, respectively. \textbf{A}) Mn off-center displacement $\delta$
along the [100] axis direction. \textbf{B}) Mn off-center displacement $\delta$
along the [110] axis direction. For tensile strain, we focus on Mn off-center displacement $\delta$
along the [110] axis direction.}
\end{figure}

In this sub-section, we study perovskite
Sr$_{0.5}$Ba$_{0.5}$MnO$_3$ alloy under epitaxial strain.
To simulate cation alloys, we employ a simulation cell with a checkerboard
arrangement of Sr-Ba cations (see Fig.~\ref{fig:SBMOstructure}).

\begin{figure}[t]
\includegraphics[angle=0,width=0.9\textwidth]{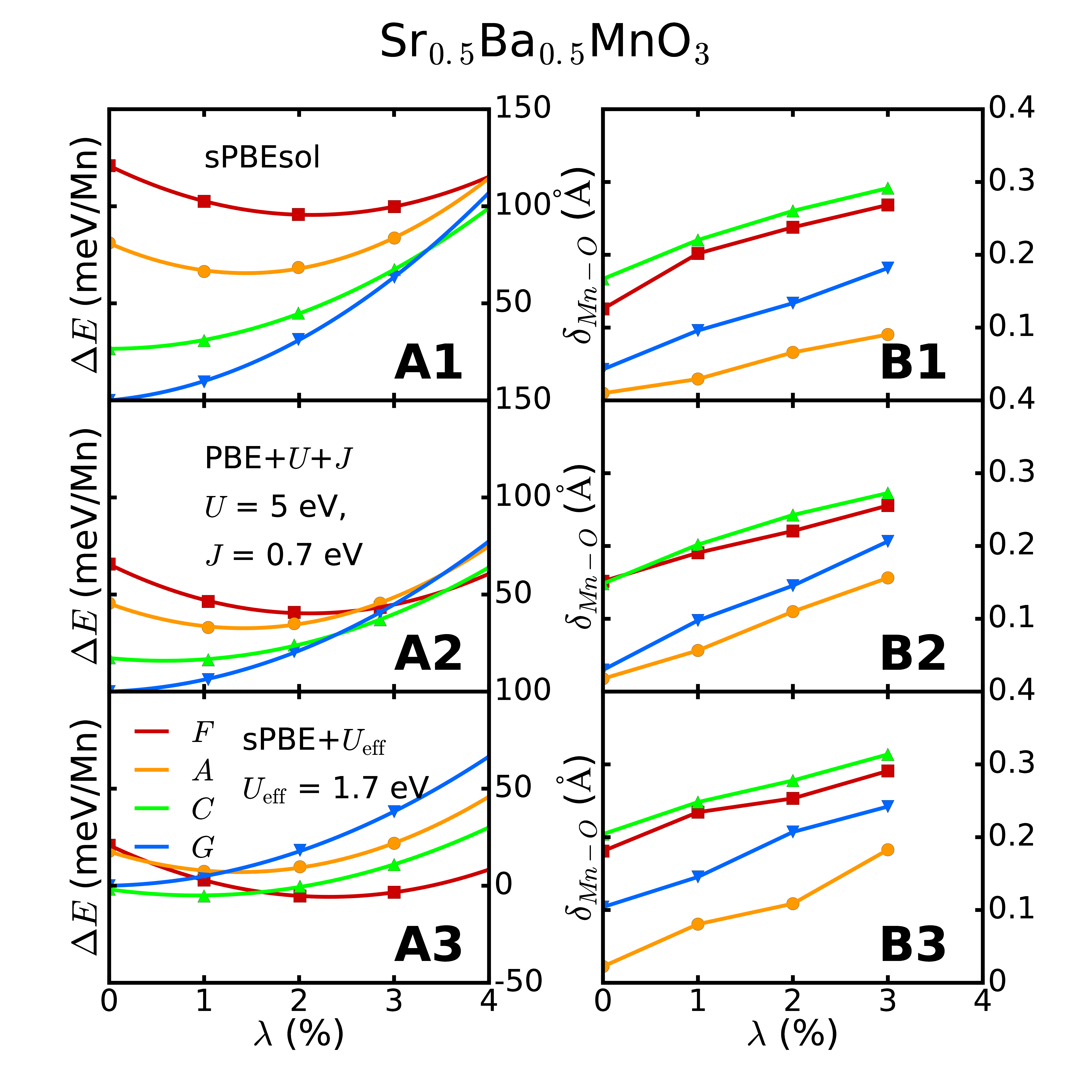}
\caption{\label{fig:strain_SBMO} Perovskite
  Sr$_{0.5}$Ba$_{0.5}$MnO$_3$: \textbf{A1}) and \textbf{B1}) using the
  sPBEsol method.  \textbf{A1}) Total energy
  for different magnetic states under tensile strain. The zero point
  is chosen as the total energy of $G$-type antiferromagnetic state at
  zero strain. $\lambda$ is the strain with respect to the equilibrium
  lattice constant of $G$-type antiferromagnetic state. \textbf{B1})
  Mn off-center displacement along the [110] direction of SrMnO$_3$
  for different magnetic states under tensile strain.  \textbf{A2})
  and \textbf{B2}) using the PBE+$U$+$J$ method with $U$ = 5 eV and
  $J$ = 0.7 eV.  \textbf{A2}) same as \textbf{A1}).  \textbf{B2}) same
  as \textbf{B1}).  \textbf{A3}) and \textbf{B3}) using the
  sPBE+$U_{\rm eff}$ method with $U_{\rm eff}$ = $U-J$ = 1.7 eV.
  \textbf{A3}) same as \textbf{A1}).  \textbf{B3}) same as
  \textbf{B1}).  The red squares and lines are the ferromagnetic state
  ($F$). The orange squares and lines are the $A$-type
  antiferromagnetic state ($A$). The green squares and lines are the
  $C$-type antiferromagnetic state ($C$). The blue squares and lines
  are the $G$-type antiferromagnetic state ($G$).}
\end{figure}

To be consistent with the study of strained SrMnO$_3$ in the previous
sub-section, we consider the Mn off-center displacement along the
[110] direction and apply tensile strain, which can enhance the [110]
polarization. Following the same convention, the biaxial strain
$\lambda$ is defined in Eq.~(\ref{eqn:lambda}). Like SrMnO$_3$, we
also study four common magnetic orderings ($F$, $A$, $C$, $G$) in
Sr$_{0.5}$Ba$_{0.5}$MnO$_3$ alloy. The panels \textbf{A} of
Fig.~\ref{fig:strain_SBMO} present the total energy of different
magnetic orderings of the Sr$_{0.5}$Ba$_{0.5}$MnO$_3$ alloy as a
function of tensile strain, using sPBEsol, PBE+$U$+$J$ and
sPBE+$U_{\rm eff}$. First we notice that for all three methods, at
$\lambda$ = 0, the $G$-type antiferromagnetic state has lower energy
than the ferromagnetic state. With an increasing lattice constant $a$,
the total energy of the $G$-type antiferromagnetic ordering
monotonically increases, while the total energy of the ferromagnetic
state first decreases to the minimum at its own equilibrium position
and then increases with strain. In particular, we notice that for the
sPBE+$U_{\rm eff}$ method, the minimum of the ferromagnetic state is
lower than the minimum of the $G$-type antiferromagnetic state. This
means that without any epitaxial strain, sPBE+$U_{\rm eff}$ predicts
that the ground state of the Sr$_{0.5}$Ba$_{0.5}$MnO$_3$ alloy is
ferromagnetic. However, sPBEsol and PBE+$U$+$J$ predict that the
ground state of the Sr$_{0.5}$Ba$_{0.5}$MnO$_3$ alloy is $G$-type
antiferromagnetic. Therefore, using sPBEsol and PBE+$U$+$J$ methods,
we predict that we need a critical tensile strain (5.2\% and 3.5\%,
respectively) to stabilize the ferromagnetic state.

For all three methods, the critical strain is predicted to be
smaller than that of pure SrMnO$_3$ because the presence of Ba
increases the Mn off-center displacement and favors the ferromagnetic
ordering~\cite{Lee-PRL-2010}, which acts as an ``effective strain''.
The panels \textbf{B} of Fig.~\ref{fig:strain_SBMO} show the Mn off-center displacement
$\delta_{\textrm{Mn-O}}$, which monotonically increases as a function
of strain. For each method, at the critical strain, the Mn off-center displacement
$\delta_{\textrm{Mn-O}}$ of the ferromagnetic state exceeds 0.25~\AA,
which is larger than the Ti off-center displacements of commercially
important ferroelectrics BaTiO$_3$.

\subsection{Perovskite BaMnO$_3$}

\begin{figure}[t]
\includegraphics[angle=0,width=0.9\textwidth]{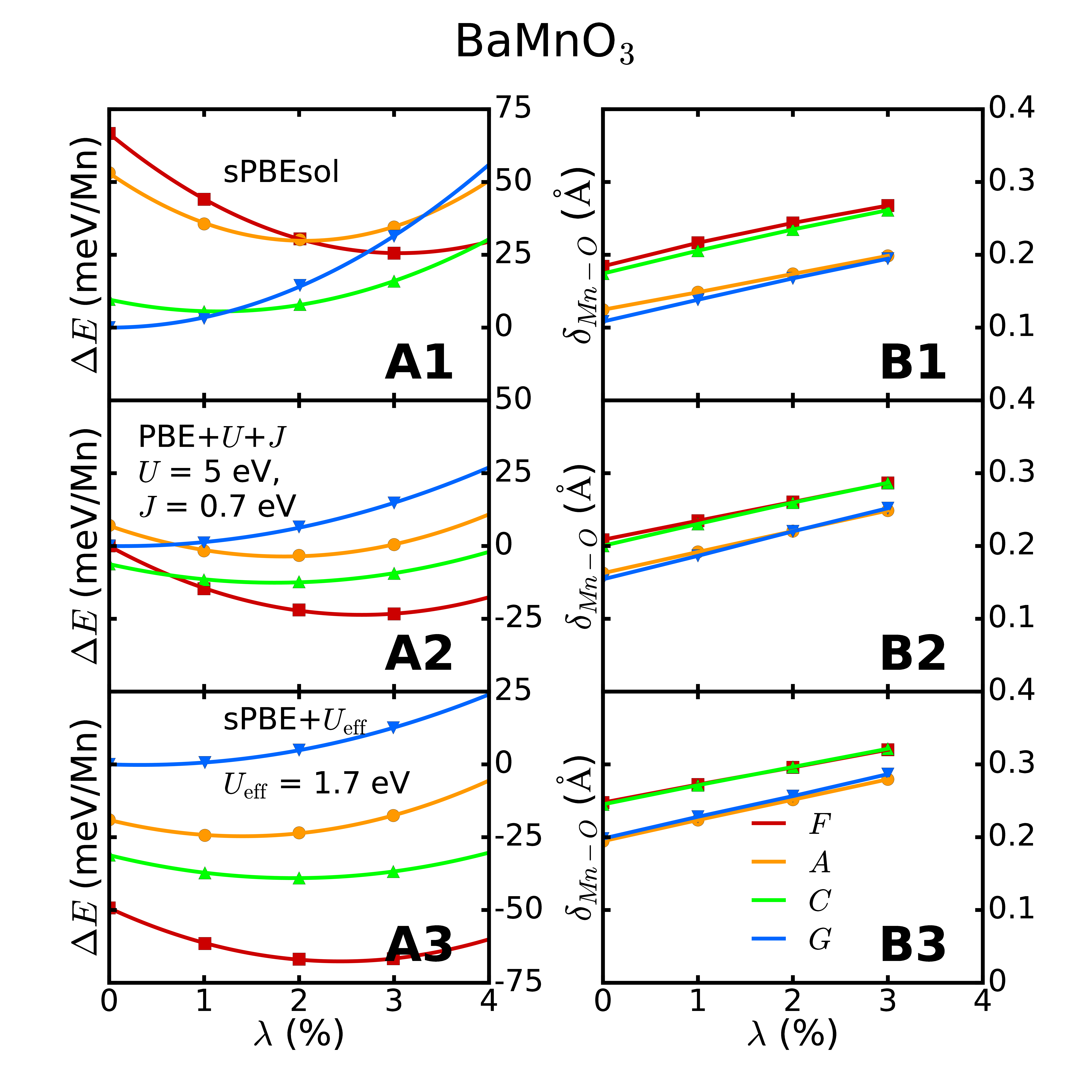}
\caption{\label{fig:strainBMO} Perovskite BaMnO$_3$: \textbf{A1}) and \textbf{B1}) using the
  sPBEsol method.  \textbf{A1}) Total energy 
  for different magnetic states under tensile strain. The
  zero point is chosen as the total energy of $G$-type
  antiferromagnetic state at zero strain. $\lambda$ is the strain with
  respect to the equilibrium lattice constant of $G$-type
  antiferromagnetic state. \textbf{B1}) Mn off-center displacement
  along the [110] direction of SrMnO$_3$ for different magnetic states
  under tensile strain.  \textbf{A2}) and \textbf{B2}) using the
  PBE+$U$+$J$ method with $U$ = 5 eV and $J$ = 0.7 eV.  \textbf{A2})
  same as \textbf{A1}).  \textbf{B2}) same as \textbf{B1}).
  \textbf{A3}) and \textbf{B3}) using the sPBE+$U_{\rm eff}$ method
  with $U_{\rm eff}$ = $U-J$ = 1.7 eV.  \textbf{A3}) same as
  \textbf{A1}).  \textbf{B3}) same as \textbf{B1}).  The red squares
  and lines are the ferromagnetic state ($F$). The orange squares and
  lines are the $A$-type antiferromagnetic state ($A$). The green
  squares and lines are the $C$-type antiferromagnetic state ($C$). The
  blue squares and lines are the $G$-type antiferromagnetic state ($G$).}
\end{figure}

In this sub-section, we study perovskite BaMnO$_3$ under epitaxial strain.
The simulation cell is shown in Fig.~\ref{fig:BMOstructure}. We follow
the convention of the previous two sub-sections. The results are shown
in Fig.~\ref{fig:strainBMO}. From the panels \textbf{A} of
Fig.~\ref{fig:strainBMO}, both PBE+$U$+$J$ and sPBE+$U_{\rm eff}$
predict that without any epitaxial strain, the ground state of
perovskite BaMnO$_3$ is ferromagnetic. However, sPBEsol predict that
the ground state of perovskite BaMnO$_3$ is still $G$-type
antiferromagnetic ordering and 4\% tensile strain is needed to
stabilize the ferromagnetic state. From the panels \textbf{B} of Fig.~\ref{fig:strainBMO}, we find that
within a 4\% tensile strain, all the three methods predict a robust Mn
off-center displacement for the four magnetic orderings investigated.

Finally, we summarize the critical tensile strain for
Sr$_{1-x}$Ba$_{x}$MnO$_3$ ($x$ = 0, 0.5 and 1) calculated with
different exchange correlation functionals. It is evident that
for all three methods used here, increasing Ba doping decreases the critical
strain that stabilizes the ferromagnetic ordering. For each Ba doping
$x$ in Sr$_{1-x}$Ba$_x$MnO$_3$, sPBE+$U_{\rm eff}$ most favors
ferromagnetism, while sPBEsol least favors ferromagnetism. This is
consistent with Fig.~\ref{fig:energydiffdoping}. 

We also studied the compressive strain (not shown here), which enhances
the polarization along the [100] direction. For compressive strain,
the magnetic energy dependence on strain and exchange correlation
functional is very similar to what we have found for tensile strain.

\begin{table}[h!]
\caption{\label{tab:critical} Critical tensile strain that stabilizes
    ferromagnetic state in Sr$_{1-x}$Ba$_x$MnO$_3$ ($0 \leq x \leq 1$),
    calculated using the sPBEsol, PBE+$U$+$J$ and sPBE+$U_{\rm eff}$
    methods. In the PBE+$U$+$J$ method, we use
    $U$ = 5 eV and $J$ = 0.7 eV. In the sPBE+$U_{\rm eff}$ method, we use
    $U_{\rm eff}$ = 1.7 eV.} 
\begin{center}
\begin{tabularx}{\textwidth}{c *{3}{|Y} c}
\hline\hline
  &   sPBEsol  & PBE+$U$+$J$  & sPBE+$U_{\rm eff}$\\
\hline
SrMnO$_3$   &   6.8\%    &   4.8\%      & 3.2\%  \\
Sr$_{0.5}$Ba$_{0.5}$MnO$_3$    &  5.2\%   &  3.5\%  & no strain needed \\
BaMnO$_3$   &  4.0\%  & no strain needed &  no strain needed \\
\hline\hline
\end{tabularx}
\end{center}
\end{table}

\section{\label{sec:pressure} Pressure stabilization of perovskite B\lowercase{a}M\lowercase{n}O$_3$}

While our calculations predict that perovskite
Sr$_{1-x}$Ba$_{x}$MnO$_3$ has appealing multiferroic properties, a
hexagonal polymorph of manganites has lower energy than the perovskite
structure ~\cite{Sondena-PRB-2006, Sondena-PRB-2006b}. However, the
perovskite polymorph may be stabilized by
pressure~\cite{Nielsen-PRB-2014}. In this section, we focus on
BaMnO$_3$ and study the energetics of different polymorphs as a
function of pressure. All the results presented are obtained using
the PBE+$U$+$J$ method ($U$ = 5 eV and $J$ = 0.7 eV). Other methods,
such as sPBEsol and sPBE, yield qualitatively consistent conclusions,
but with a critical pressure 20\% larger in magnitude (see below).

\begin{figure}[t]
\includegraphics[angle=0,width=\textwidth]{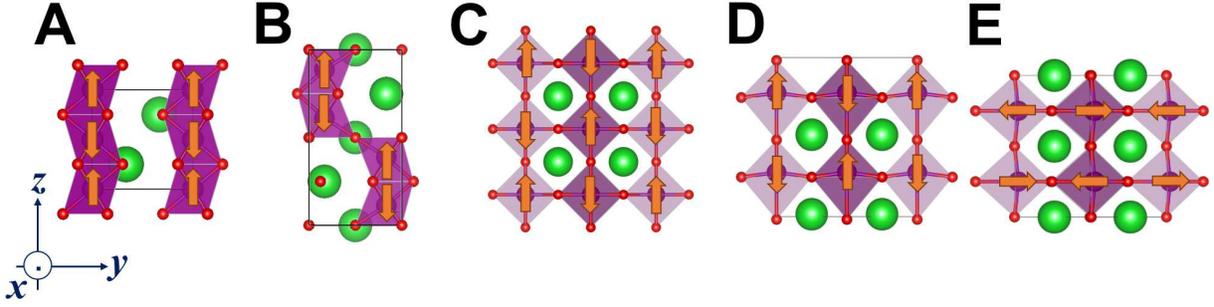}
\caption{\label{fig:fullstructure} Five different atomic structures of
  BaMnO$_3$. The green balls are La atoms. The purple cages are
  MnO$_6$.  \textbf{A}) the hexagonal structure with face-sharing
  MnO$_6$ oxygen octahedra; \textbf{B}) the hexagonal structure with
  mixed face-sharing and corner sharing MnO$_6$ oxygen octahedra;
  \textbf{C}) the cubic perovskite structure with corner-sharing
  MnO$_6$ oxygen octahedra; \textbf{D}) the tetragonal perovskite
  structure with Mn off-center displacement along the $z$ direction;
  \textbf{E}) the orthorhombic perovskite structure with Mn off-center
  displacement in the $xy$ plane. The orange arrows indicate
  spin arrangements in the antiferromagnetic state considered for each
  structure.}
\end{figure}

Fig.~\ref{fig:fullstructure} shows the atomic structure of five
different polymorphs of
BaMnO$_3$. Fig.~\ref{fig:fullstructure}\textbf{A} shows the structure
of hexagonal BaMnO$_3$ with face-sharing MnO$_6$ oxygen
octahedra. This structure is called
$2H$. Fig.~\ref{fig:fullstructure}\textbf{B} is also a hexagonal
structure of BaMnO$_3$ with mixed face-sharing and corner-sharing
MnO$_6$ oxygen octahedra. This structure is called
$4H$. Fig.~\ref{fig:fullstructure}\textbf{C} is the cubic perovskite
structure of BaMnO$_3$ with corner-sharing MnO$_6$ octahedra.
This structure is referred to as ``cubic''.
Fig.~\ref{fig:fullstructure}\textbf{D} is a tetragonal perovskite
structure of BaMnO$_3$ with Mn off-center displacements along the $z$
direction. This structure is referred to as ``tetragonal''.
Fig.~\ref{fig:fullstructure}\textbf{E} is an orthorhombic perovskite
structure of BaMnO$_3$ with Mn off-center displacements in the $xy$ plane. 
This structure is referred to as ``orthorhombic''. We first
use the PBE+$U$+$J$ method to calculate the total energies of these
structures at zero pressure, which are presented in
Table~\ref{tab:difference}. We consider ferromagnetic and
antiferromagnetic states (the spin arrangement for the
antiferromagnetic state for each structure is explicitly shown as
orange arrows in Fig.~\ref{fig:fullstructure}). The zero of energy is chosen as the
energy of the antiferromagnetic state of the $2H$ structure, which has
the lowest total energy among all the structures considered. We find
that at ambient pressure all the perovskite structures of BaMnO$_3$
are about 0.7 eV/Mn higher in energy that the $2H$ structure.  For
each structure, we also calculate the ferromagnetic-antiferromagnetic
energy difference $\Delta E = E(AF) - E(F)$. We see that without Mn
off-center displacement ($2H$, $4H$ and cubic structures), the
antiferromagnetic state has lower energy than the ferromagnetic state.
With Mn off-center displacements (tetragonal and orthorhombic
structures), the ferromagnetic state has lower energy. This further
indicates that Mn off-center displacements help stabilize ferromagnetism. We
also calculate the volume per Mn atom of each structure. We notice
that antiferromagnetic state of the cubic structure has the smallest
volume 61.9~\AA$^3$ per Mn, while the antiferromagnetic $2H$ structure
that has the lowest total energy has a much larger volume
71.0~\AA$^3$. Face-sharing oxygen octahedra are in fact more closely
packed than corner-sharing oxygen octahedra. However, in the $2H$
structure, between the ``column'' of face-sharing oxygen octahedra 
there is hollow space, which increases the volume.

\begin{table}[t!]
\caption{\label{tab:difference} Total energies $E$ with the zero point
  chosen as the total energy of $2H$ structure with antiferromagnetic
  ordering, $\Delta E = E(AF) - E(F)$ the energy difference between
  ferromagnetic ordering and antiferromagnetic ordering for each
  structure, and equilibrium volume $\Omega$ per Mn atom, calculated
  using the PBE+$U$+$J$ method with $U$ = 5 eV and $J$ = 0.7 eV.}
\begin{center}
\begin{tabularx}{\textwidth}{c *{10}{|Y} c}
\hline\hline
structure & \multicolumn{2}{c|}{2$H$} & \multicolumn{2}{c|}{4$H$} & \multicolumn{2}{c|}{cubic} & \multicolumn{2}{c|}{tetragonal} & \multicolumn{2}{c}{orthorhombic}\\
\hline
ordering & FM & AFM & FM & AFM & FM & AFM & FM & AFM & FM & AFM\\
\hline
$E$ (meV/Mn) &  103  & 0  & 186  & 109  & 837 & 775 & 724 & 746 & 727 & 754 \\
\hline
$\Delta E$ (meV/Mn) & \multicolumn{2}{c|}{-103} & \multicolumn{2}{c|}{-77} &
\multicolumn{2}{c|}{-62} & \multicolumn{2}{c|}{22} & \multicolumn{2}{c}{27}\\
\hline
volume $\Omega$ (\AA$^3$/Mn) & 71.6 & 71.0 & 66.4 & 66.1 & 62.9 & 61.9 & 65.9 & 64.1 & 65.8 & 63.4\\
\hline\hline
\end{tabularx}
\end{center}
\end{table}

\begin{figure}[t]
\includegraphics[angle=0,width=0.85\textwidth]{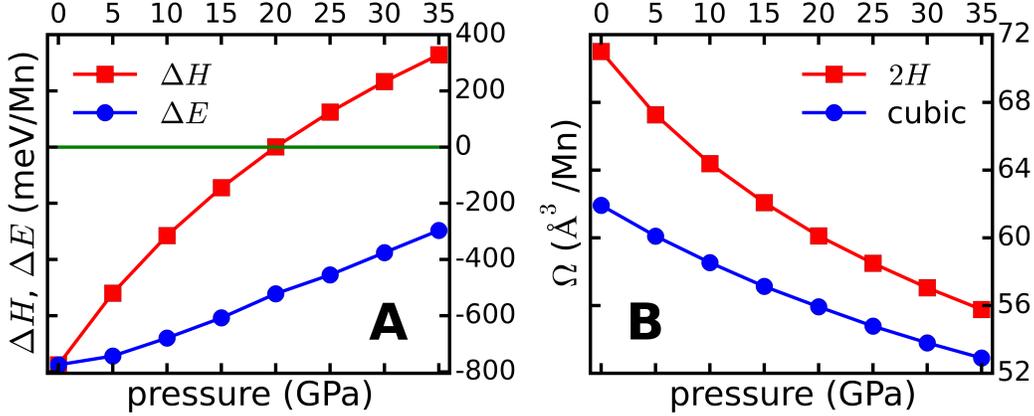}
\caption{\label{fig:pressure} Pressure dependence of the $2H$
structure and the
  cubic structure. For the $2H$ structures, we study the
  antiferromagnetic state; for the cubic structure, we study the
  $G$-type antiferromagnetic state.  The results are calculated
  using the PBE+$U$+$J$ method with $U$ = 5 eV and $J$ = 0.7 eV. \textbf{A})
  Enthalpy difference between the $2H$ structure and the cubic
  structure $\Delta H = H(2H) - H (\textrm{cubic})$ (red
  curves). Energy difference between the $2H$ structure and the cubic
  structure $\Delta E = E(2H) - E (\textrm{cubic})$ (blue curves). The
  green line highlights the structural transition point.
  \textbf{B}) Volume $\Omega$ per Mn atom of the $2H$ structure (red curves)
  and the cubic structure (blue curves). }
\end{figure}

The volume comparison suggests that external pressure can stabilize
the cubic perovskite structure over the hexagonal $2H$ structure as
found experimentally~\cite{Sondena-PRB-2006}.  To test this
hypothesis, we calculate the total enthalpy as well as total energy of
the antiferromagnetic $2H$ structure and antiferromagnetic cubic
structure. The enthalpy difference
$\Delta H = H(2H) - H(\textrm{cubic})$ and the energy difference
$\Delta E = E(2H) - E(\textrm{cubic})$ are shown in
Fig.~\ref{fig:pressure}\textbf{A}. We find that at a critical pressure
of 20 GPa, there is a transition from the $2H$ structure to the cubic
structure.  Compared to the energy difference $\Delta E$, we can
clearly see that it is the volume difference
$\Delta V = V(2H) - V(\textrm{cubic})$ that stabilizes the cubic
structure via pressure.  A 20 GPa pressure is experimentally
achievable and has been applied on cubic
SrMnO$_3$~\cite{Nielsen-PRB-2014}.  To make the discussion complete,
we present in Fig.~\ref{fig:pressure}\textbf{B} the volume per Mn of
the $2H$ structure and of the cubic structure as a function of
pressure. We see that because the $2H$ structure is more hollow than
the cubic structure, it is more compressible (the volume decreases
faster with pressure).  However, up to the critical pressure, the
volume of the $2H$ structure is always larger than that of the cubic
structure, which results in the pressure-driven structural
transition. We need a few comments here: i) we verified that with the
application of pressure, there is no magnetic transition for the $2H$
structure and the cubic structure, i.e. the
antiferromagnetic ordering always has lower energy than the
ferromagnetic ordering; ii) at the critical pressure, the Mn
off-center displacements are suppressed, i.e.  the tetragonal and
orthorhombic perovskite structures are reduced to the cubic structure;
iii) we repeated all the calculations using the sPBEsol and sPBE methods
and find all the results are qualitatively consistent with a critical
pressure of 21 GPa for sPBEsol and 25 GPa for sPBE.

\section{\label{sec:conclusion} Conclusions}

We have systematically studied the phase diagram of perovskite
Sr$_{1-x}$Ba$_x$MnO$_3$ ($0\leq x \leq 1$) as a function of Ba
concentration $x$ and epitaxial strain, as well as the phase diagram
of hexagonal polymorphs of
BaMnO$_3$ as a function of external pressure.

For perovskite Sr$_{1-x}$Ba$_x$MnO$_3$, we find that both increasing
the Ba concentration and imposing epitaxial strain tend to stabilize a
ferromagnetic-ferroelectric state. In the vicinity of the phase
boundary between a ferromagnetic state with a larger polarization and
an antiferromagnetic state with a smaller polarization, either
applying a magnetic field or an electric field
can induce a first-order phase transition in polarization or magnetization, and thus realize an
effective giant magneto-electric coupling.  However, the details of
the phase diagram, in particular the critical Ba doping and critical
epitaxial strain strongly depend on the choice of exchange correlation
functional. We find that among the three flavors of PBE exchange
correlation functionals (sPBEsol, PBE+$U$+$J$ and
sPBE+$U_{\rm eff}$), sPBEsol least favors the ferromagnetism and
predicts the largest critical Ba concentration and largest critical strain,
while sPBE+$U_{\rm eff}$ most favors the ferromagnetism and therefore
predicts the smallest critical Ba concentration and smallest critical epitaxial
strain (above some Ba concentration, epitaxial strain is not even
needed to stabilize the ferromagnetic-ferroelectric state).  We also
show that a high pressure of over 20 GPa can stabilize the perovskite
structure of BaMnO$_3$ over its hexagonal polymorphs.

Perovskite Sr$_{1-x}$Ba$_x$MnO$_3$ with $x > 0.5$ has not yet been 
synthesized in bulk or grown in thin film form~\cite{Langenberg-ACSAMI-2015}. 
We hope that our theoretical predictions on multiferroic properties 
of perovskite Sr$_{1-x}$Ba$_x$MnO$_3$ could stimulate further experiments. 

\begin{acknowledgments}
H. Chen is supported by National Science Foundation under Grant
No. DMR-1120296.  A. J. Millis is supported by National Science
Foundation under grant No. DMR-1308236. A. J. Millis thanks the
College de France for hospitality and a stimulating intellectual
environment while this paper was being prepared.
\end{acknowledgments}


\end{document}